\documentclass[aps,prl,twocolumn,superscriptaddress]{revtex4-2}

\usepackage{tikz}
\usetikzlibrary{shapes.geometric, arrows}
\usepackage{graphicx}
\usepackage{amsmath}
\usepackage{color}
\usepackage{gensymb}
\usepackage{lineno}
\usepackage[caption=false]{subfig}

\begin{document}



\title{Measurement of low energy nuclear recoil events with the Phonon-mediated voltage-assisted Hybrid Detector for rare event searches}


\author{S. Maludze}
\author{M. Mirzakhani}
\author{W. Baker}
\author{M. Lee}
\author{C. Savage}
\author{H. Neog}
\author{R. Mahapatra}
\author{N. Mirabolfathi}
\author{M. Platt}
\author{A. Jastram}

\affiliation{Department of Physics and Astronomy, Texas A\&M University, College Station, TX, 77843}


\date{\today}

\begin{abstract}
The phonon-mediated hybrid detector is made out of a monolithic silicon crystal characterized by two interconnected regions linked through a narrow neck. Operating solely on phonon signal measurements, the hybrid design facilitates the differentiation between electron recoil and nuclear recoil events, effectively discerning two types of interaction down to low energy levels. With a newly implemented software triggering technique (SWT), low-energy nuclear recoil events of approximately $500~eV_{ee}$ have been measured. In addition to this, an electron recoil background reduction of up to $95\%$ has been successfully demonstrated.

\end{abstract}


\maketitle


\section{Introduction}
Numerous astronomical observations have presented evidence suggesting that the predominant constituent of the Universe's matter content is non-luminous and non-baryonic in nature, commonly referred to as dark matter \cite{Zwicky,Rubin1}. At the same time, Only $~5\%$ of the universe is composed of observable matter.  Among the leading contenders for the identity of dark matter is the Weakly Interacting Massive Particle (WIMP), characterized by its interaction on the weak scale \cite{WIMP, WIMP_exp}. Various experiments  \cite{SuperCDMS, DAMIC, SENSEI,CRESST-III} employ direct detection methodologies, wherein the interaction between dark matter and normal matter is probed through sensitive detector technologies and background mitigation techniques. To probe the parameter space at lower mass levels than the current detection thresholds allow, it becomes imperative to possess detectors featuring lower thresholds and substantial target masses.

In the conventional configuration of a Cryogenic Dark Matter Search (CDMS) apparatus, the iZIP (interleaved Z-sensitive Ionization and Phonon) detector is equipped with both phonon sensors and charge electrodes, enabling the simultaneous measurement of phonon and ionization signals \cite{Agnese2013_2}. This two-signal detection serves as a tool for identifying a type of particle interaction within the detector, specifically distinguishing electron recoil (ER) and nuclear recoil (NR) events. Given the greater ionization yield for ER events, a fraction of two signals is used as a discriminator factor. In contrast, a high-voltage cylindrical detector, characterized by a lower threshold compared to iZIP, is equipped solely with phonon sensors and lacks this discrimination capability \cite{cdmslite}. 

A novel hybrid detector has been developed with the motivation to combine the advantageous features of these two technologies \cite{neog}. This innovation employs a monolithic silicon crystal characterized by two distinct regions: a low-voltage domain with a substantial volume and a high-voltage domain with a relatively smaller volume. Those two regions are linked by a narrow segment facilitating the propagation of charge carriers to the high-voltage region.

Within this hybrid detector, the low-voltage region constitutes the fiducial volume where particle interactions occur. Phonon sensors, positioned at the base of the low-voltage region, capture phonon signals generated at the interaction point. On the other hand, electrons produced during the interaction are guided toward the high-voltage region due to an applied electric field. While traversing the high-voltage domain, these electrons generate more Neganov-Trofimov-Luke (NTL) phonons \cite{Neganov, Luke} with a total signal proportional to the applied voltage and the number of electron-hole pairs generated as a result of an initial interaction. These NTL phonons are then collected by sensors located at the top of the high-voltage region. As a result of this setup, the hybrid detector, equipped solely with phonon sensors, possesses the remarkable capability to capture and quantify both phonon (lower region) and ionization (upper region) signals. The fraction of these signals provides a robust discriminator for distinguishing between ER and NR events, thereby enhancing the detector's capacity for background rejection on an event-by-event basis.

This paper presents the updated results of the hybrid performance with a newly developed software triggering technique (SWT). With this technique, it has been possible to demonstrate the hybrid detector's remarkable capability to distinguish between ER and NR down to low energies, which represents the region of interest (ROI) for rare event searches. 

\section{Working principle}
The hybrid detector consists of a monolithic silicon crystal and has two regions separated by a narrow neck. The lower side of the crystal has a truncated conical geometry with dimensions of 1.4 cm in height, 7.6 cm in bottom diameter, and 4.8 cm in top diameter. The conical shape pushes electric field lines toward the high-voltage region inside the crystal. The upper side takes the form of a small cylinder with a diameter of 2.5 cm and a height of 1 cm. A narrow 0.5 mm gap separates these two regions. The detector's fiducial mass (mass of the low-voltage side) is $230~g$, while the total mass is $250~g$. 

\begin{figure}[h!]
\centering
\includegraphics[width=0.9\linewidth, height=0.7\linewidth]{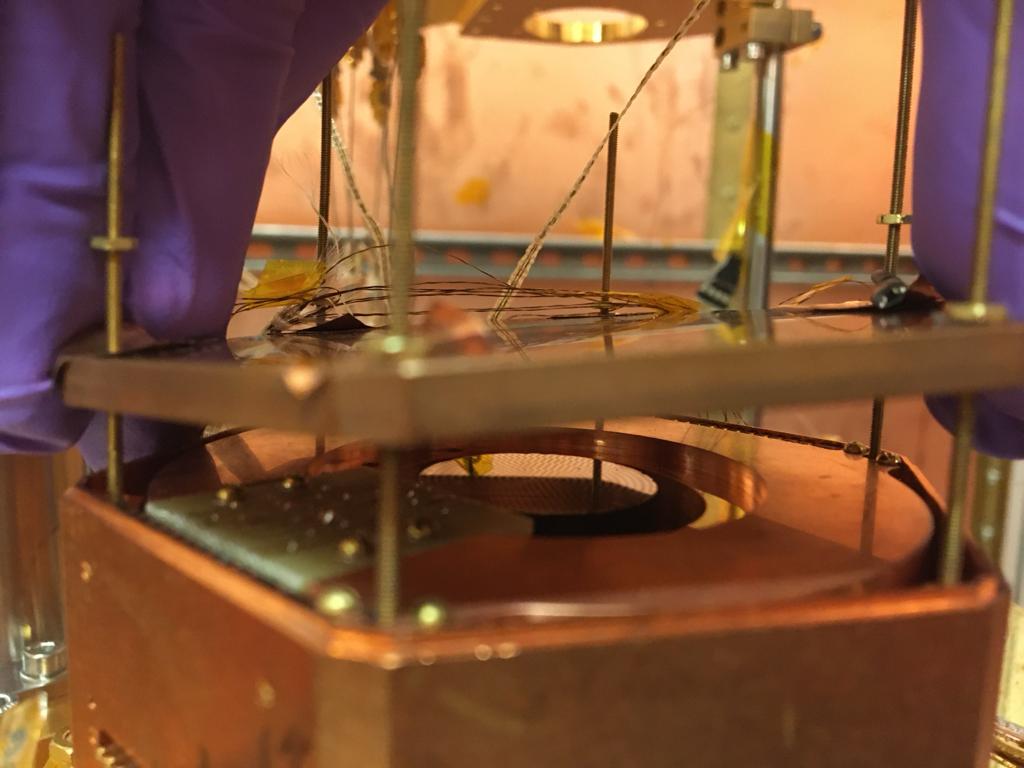}
\caption{Picture of the top side of the Hybrid detector in the experimental setup.}

\label{Hybrid_picture}
\end{figure}

On the crystal's top and bottom surfaces, phonon Transition-Edge-Sensors (TESs) \cite{QET:1995} are photolithographically patterned. The top surface has a single readout channel, referred to as a top channel, while the bottom surface is subdivided into four readout channels: three inner channels and one outer channel, denoted as the bottom channels. To ensure its proper functioning, the detector is placed in a copper housing that not only shields it from external background radiation but also plays a pivotal role in electric field shaping within the crystal (Figure \ref{Hybrid_picture}). Proper electric field shaping is essential for the operation of the detector.

When an interaction occurs within the low-voltage domain, which constitutes the fiducial volume, primary phonons, and electron-hole pairs are generated in close proximity to the interaction point. These primary phonons bounce back and forth within the low-voltage domain until their eventual absorption by the bottom channels.  Simultaneously, the electron-hole pairs drift along the electric field lines within the domain, generating NTL phonons in the low-voltage region, which are also captured by the bottom channels. Notably, the contribution of NTL phonons to the signal registered on the bottom channels is relatively lower compared to primary phonons. At the same time, the electric field lines direct electrons toward the high-voltage domain, where they also generate NTL phonons. These phonons are collected by the top channel. Thus, the signal observed on the top channel is directly proportional to the number of electrons created as a result of the initial interaction, providing an indirect measure of ionization.

A notable feature of the hybrid detector is its capability to estimate both the ionization yield (the Lindhard factor for nuclear recoil events \cite{Lindhard}) and the recoil energy based on signals from the top and bottom channels, considering that the total phonon signal has a contribution from both sides of the crystal. Leakage factor parameters that account for this contribution, are determined experimentally from the calibration dataset (Pair of formulas presented in \cite{neog}). 

To summarize the hybrid detector working principle, the signal acquired from the bottom channels measures the contribution of primary phonons and a comparably smaller amount of NTL phonons generated in the low-voltage region. In contrast, the top channel detects ionization by capturing NTL phonons generated in the high-voltage domain. A simple ratio of the signals from the top and bottom channels denoted as $Y$, effectively carries information about the ionization yield.  

\section{Software trigger and data analysis pipeline}
In this hybrid experiment, we introduced the software triggering technique (SWT).  In previous runs, event triggering relied on manual threshold settings based on the real-time display of pulses via the digitizer software. However, this approach came with significant drawbacks. Notably, it necessitated human judgment to differentiate noise from genuine pulses, potentially leading to the oversight of low-energy events that exceeded the detector threshold by a considerable margin. Once data acquisition was completed, the trigger threshold was fixed and unmodifiable. Under such circumstances, implementing a new trigger threshold necessitated the collection of an entirely new dataset. The hardware trigger method served its purpose well in calibration runs — where calibration lines comfortably exceeded the threshold and pulse-noise differentiation was evident. 

\begin{figure}[h!]
\centering
\includegraphics[width=\linewidth, height=0.15\linewidth]{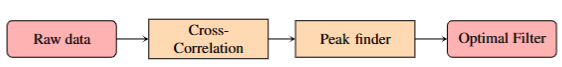}
\caption{Flaw chart illustrating steps for updated analysis pipeline with SWT technique.}

\label{Flow_chart}
\end{figure}

With the SWT method, the signal from the detector is continuously recorded. After recording the data in a desired format, individual pulses are found by the software and saved in a different directory. The first step in the process is Cross-Correlation - a pre-selected pulse template is cross-correlated to the entire readout signal. This step significantly reduces the noise level and facilitates pulse finding. The next step is finding pulses in the data with the peak finder algorithm. The algorithm selects pulses with a certain prominence. Although this method has an extra step of pulse finding and takes more space and computational resources, the trigger threshold becomes a tunable parameter and could be adjusted as needed once the raw data has been taken. After individual pulses are identified and saved, the Optimal Filter (OF) \cite{OF1, OF2} algorithm is implemented for event reconstruction. The algorithm takes noise and pulse templates generated from the data and makes a fit to the pulses. The primary output of this step is the amplitude of a pulse which is then converted into energy units. This is the efficient way to estimate event energy considering the noise environment.    

\begin{figure}[h!]
\centering
\includegraphics[width=\linewidth, height=0.5\linewidth]{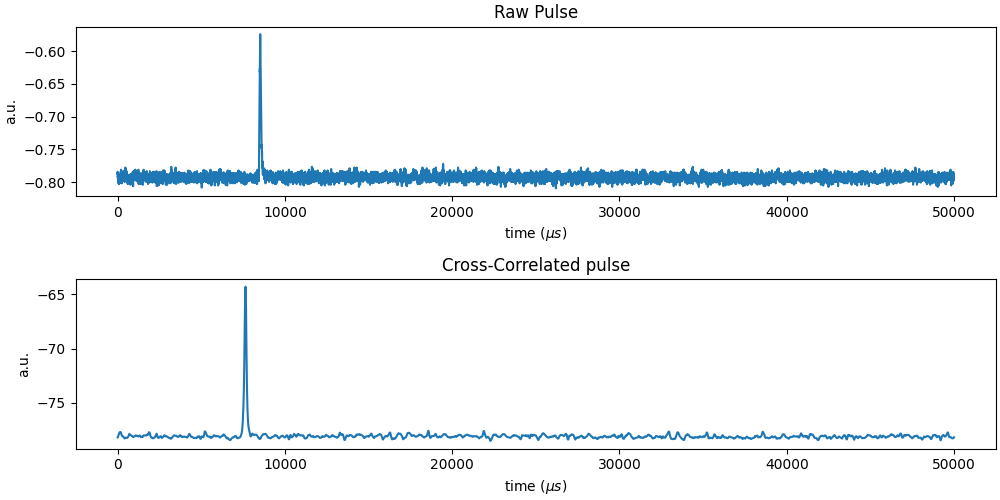}
\caption{An example of a trace (readout signal) containing a pulse and its cross-correlated version.}

\label{Flow_chart}
\end{figure}

\section{Experimental configuration and results}
As described above, the hybrid detector requires bias voltage to be applied across the crystal for its operation. 
During this experimental run, numerous datasets were acquired to ascertain the optimal signal-to-noise ratios and bias voltage settings. For more accurate electron recoil calibration, an $^{241}Am$ source \cite{AmManu} was strategically positioned at the midpoint of the bottom side of the detector. For investigating the detector's response to nuclear recoil events, $^{252}Cf$ neutron source was placed $2~ft$ away from the detector with an appropriate shielding for gamma radiation. Data was acquired with a software-triggering scheme to record near-threshold events more efficiently.   

The hybrid detector characterization process follows several steps - first, a dataset without any applied voltage across the detector is taken. With known calibration source lines, calibration factors and phonon leakage parameters are calculated.  Subsequently, the detector is biased with its optimal operation voltage to investigate its response to both electron and nuclear recoil events for an extended period.

A detector operating under the applied voltage experiences a phenomenon known as charge build-up, wherein charge carriers become trapped within crystal impurities, gradually diminishing the overall effective electric field over time. To alleviate the accumulation of trapped charge carriers and restore the detector's operational state, Light-Emitting Diodes (LEDs) are turned on periodically for a brief moment of time (LED flashing). Following the activation of an LED, a cooling period is required for the detector to return to its equilibrium temperature, during which the data acquisition process is stopped. Repeated steps of LED flashing, stopping, and resuming the data acquisition process have been automated. In one cycle, after $20~mins$ of continuous data taking, data collection is momentarily halted, and the LED is activated for a duration of $1~s$. Subsequently, the data acquisition resumes once the detector returns to its designated operational temperature, which usually takes another $20~mins$. 

Using this experimental configuration, a combined dataset of  $\sim50~hrs$ featuring $^{252}Cf$ source was collected. The source was positioned at an approximate distance of $2~ft$ from the detector. It was shielded by a layer of 4" lead bricks, blocking the excess of gamma radiation. The detector was operated under the optimal potential difference of $16~V$ across the detector. 

\begin{figure}[h!]
\centering
\includegraphics[width=\linewidth, height=0.8\linewidth]{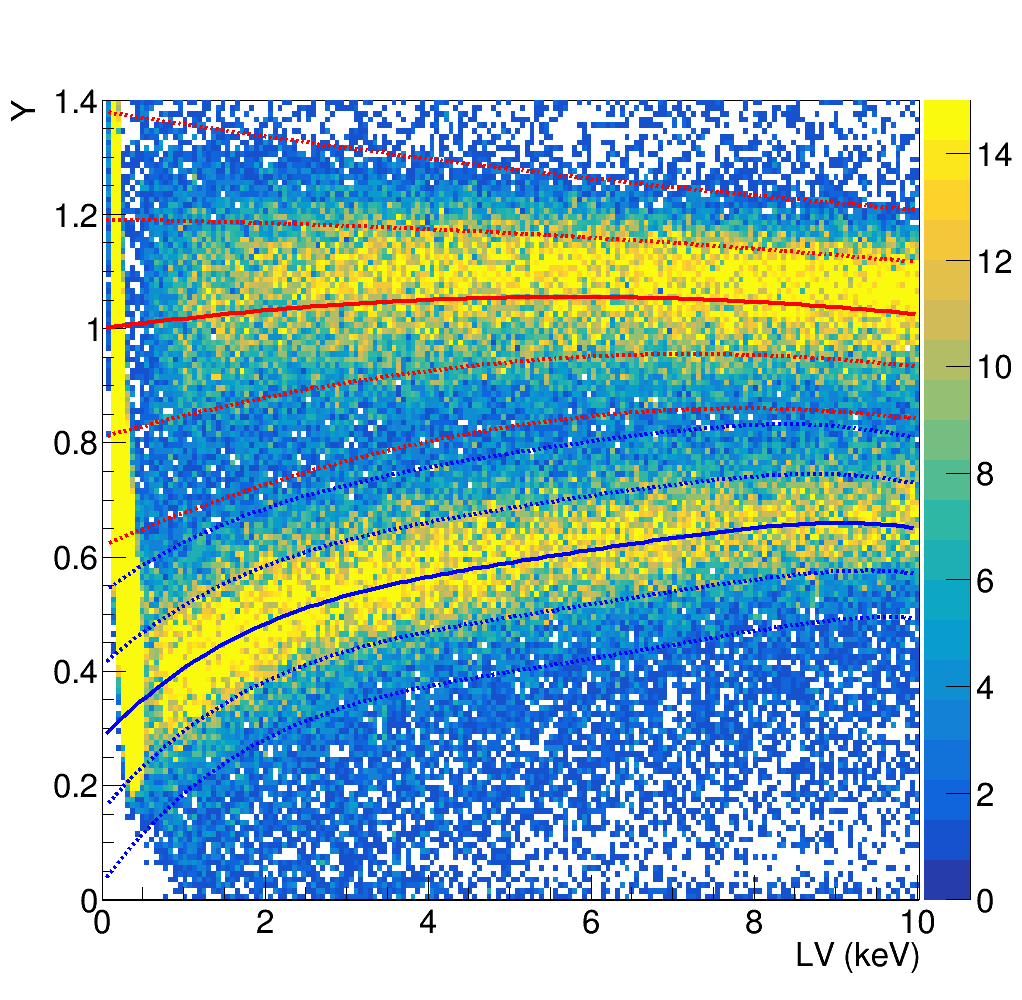}
\caption{A plot of Y against the signal on the low-voltage side from 0 to 10 keV, featuring the mean along with $1\sigma$ and $2\sigma$ of the fitted bands for electron recoil (red) and nuclear recoil (blue) events.}

\label{hybrud_Y_vs_LV_10}
\end{figure}

\begin{figure}[h!]
\centering
\includegraphics[width=\linewidth, height=0.8\linewidth]{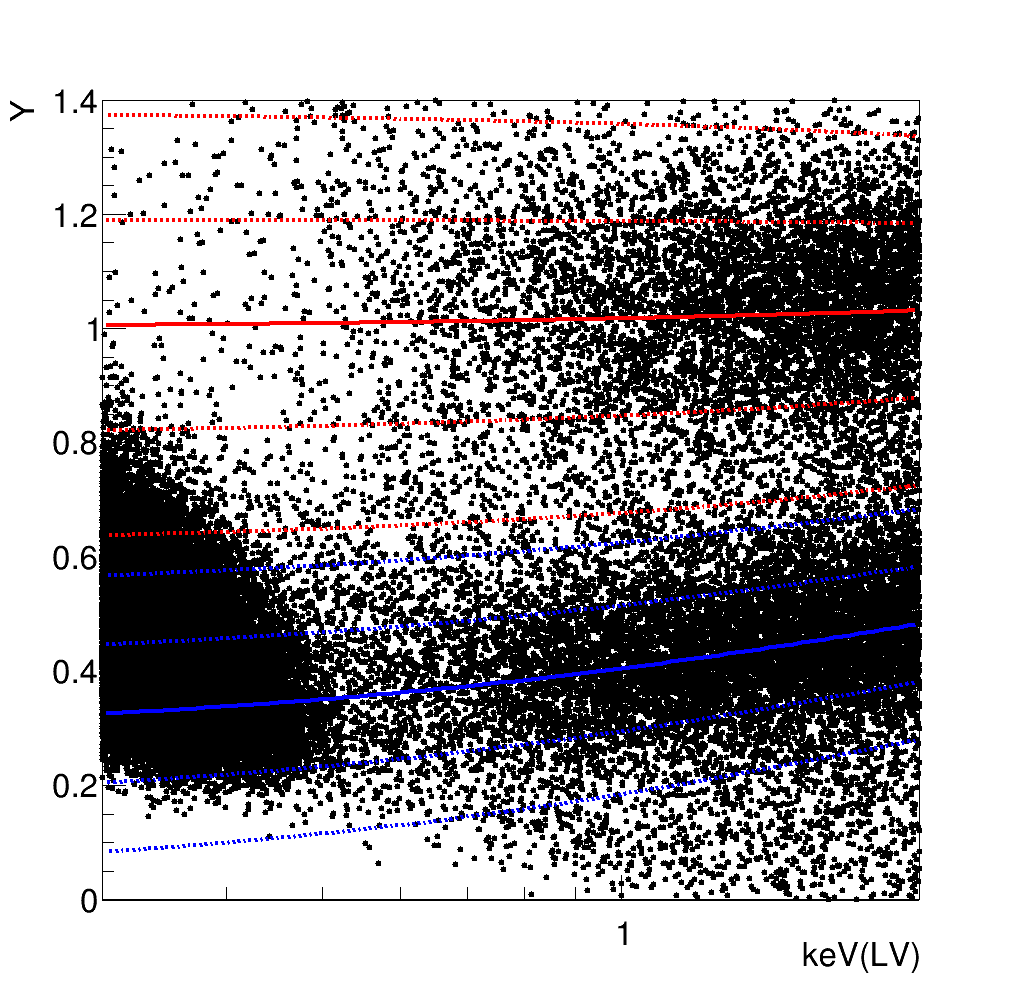}
\caption{A plot of Y against the signal on the low-voltage side from 0 to 3 keV with the same mean, $1\sigma$ and $2\sigma$ of the fitted bands. Nuclear recoil events are collected down to the energy levels of approximately $\sim500~eV_{ee}$. Red lines represent electron recoil events, while blue lines correspond to nuclear recoil events. The blob below $500~eV_{ee}$ corresponds to noise events.}

\label{hybrud_Y_vs_LV_1}
\end{figure}

Figure \ref{hybrud_Y_vs_LV_10} showcases the discriminator (Y), a fraction of the signal on the top and bottom channels, plotted against energy within the energy window of $0-10~keV$. It is important to note that the energy is represented in units of electron recoil equivalent ($keV_{ee}$). As anticipated, discernible bands corresponding to electron recoil (ER) and nuclear recoil (NR) events are evident in the figure. Additionally, the fitted mean, $1\sigma$, and $2\sigma$ bands are superimposed on the plot. Here the polynomial fit was made on energy bins with the size of $0.1~keV$. The ER event band appears uniformly distributed, while the NR band exhibits energy-dependent behavior consistent with expectations derived from the Lindhard theory \cite{Lindhard}. The primary interest of this experimental run is to assess the detector's performance and its discriminating power on near-threshold events. In Figure \ref{hybrud_Y_vs_LV_1}, energy region up to $3~keV$ is displayed (x-axis in the logarithmic scale). As we see, this detector demonstrates the capacity to measure nuclear recoil events while effectively rejecting electron recoil background down to energy levels of approximately $\sim500~eV_{ee}$.

Additionally, an extended background dataset was obtained in the absence of any source to quantify the detector's efficiency in rejecting background events. 


This estimation accounts for a fraction of the events within the NR band divided by the entire event population. 
In addition to actual NR events, it is important to note that ER events may appear within the NR band due to the incomplete charge collection. The effect of incomplete charge collection is characteristic of events happening near the conical surface/edges of the detector. The effect is caused by either not enough electric field at the peripheries to direct electrons to the neck or it can be attributed to crystal defects and impurities. With this extended source free $16~hrs$ run, on average, approximately $\sim5~\%$ of total population events appear within the NR $1\sigma$ band, leading to a substantial $95\%$ reduction in the background.

\begin{figure}[h!]
\centering
\includegraphics[width=\linewidth, height=0.8\linewidth]{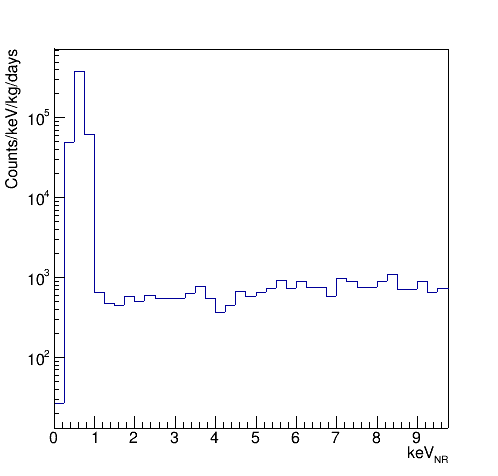}
\caption{Background run of the hybrid detector conducted throughout $16~hrs$ without any source. Approximately $3\%-10\%$ of the events manifest within the nuclear recoil band. Scaled event rate observed within the nuclear recoil band. The estimated background rate above $1~keV$ is $<900~Counts/keV/kg/days$ in the test facility without any effective shielding techniques.}

\label{Hybrid_bk-b}
\end{figure}

\section{conclusion}
Following the initial successful run that demonstrated the hybrid detector's performance, we achieved a further reduction in the energy threshold by implementing a software triggering scheme within the data processing pipeline. In addition, the detector operation under the bias voltage was automated with periodic LED flashing and cooling down, preventing significant charge build-up within the detector and enabling continuous data taking for an extended time. Consequently, we attained an energy threshold of approximately $\sim 500~eV_{ee}$, successfully demonstrating nuclear recoil/electron recoil (NR/ER) discrimination at lower energies. 
The ER band is uniform while the NR band follows the shape described by the Lindhard theory. Notably, the discrimination power improves at lower energies due to the low energy threshold. As a result, the electron recoil background reduction of $3-10\%$ has been achieved. 
Future studies will be directed toward the elimination of incomplete charge collection which would further increase the background rejection efficiency and reduce the background level for rare event searches. New designs with improved field-shaping capabilities are under consideration.




\section{acknowledgments}
This work was fully supported by DOE grants DE-SC0017859 and DE-SC0018981. We acknowledge the contribution of the key cryogenic infrastructure (Bluefors LD400) provided by NISER, India. We would like to further acknowledge the support of DAE through project Research in Basic Sciences - Dark Matter and SERB-DST through the J.C. Bose fellowship.


\bibliography{mybibfile}

\end{document}